\documentclass[a4paper,12pt]{article}
\usepackage[left=2cm,right=2cm]{geometry}
\usepackage[utf8x]{inputenc}
\usepackage{amsfonts,amsmath,amssymb,bm,eucal}
\usepackage{graphicx}
\usepackage{epstopdf}
\usepackage{subfigure}
\usepackage{color}

\title{\vspace*{-2cm}
\begin{flushright}
\normalsize{BARI-TH/2013-678}
\end{flushright}
\vspace*{1.5cm}Temperature and chemical potential dependence of the gluon condensate: a holographic study}
\author{P.~Colangelo$^a$, F.~Giannuzzi$^{a,b}$, S.~Nicotri$^{a,b}$, F.~Zuo$^a$\\~\\
\normalsize\emph{$^a$Istituto Nazionale di Fisica Nucleare, Sezione di Bari, Italy}\\
\normalsize\emph{$^b$Dipartimento di Fisica, Universit\`a degli Studi di Bari, Italy}
}
\date{}

\begin{document}

\maketitle

\begin{abstract}
The lowest dimensional gluon condensate $G_2$ is analyzed at finite temperature and chemical potential using a holographic model of QCD with conformal invariance broken by a background dilaton.
Starting from the free energy of the model, the thermodynamical quantities needed to determine the $T$ and $\mu$ dependence of the gluon condensate are evaluated.
At high temperature the gluon condensate is independent of chemical potential.
Moreover, at $\mu=0$, the temporal and spatial Wilson loops at low temperature are computed; they are  related to the (chromo) electric and magnetic components of $G_2$, respectively. The $T$-dependence of the two components is separately determined.

\vspace*{0.3cm}
\noindent{pacs:
11.25.Tq, %Gauge/string duality
11.10.Kk, %Field theories in dimensions other than four
11.15.Tk %Other nonperturbative techniques
12.38.Lg %Other nonperturbative calculations
}
\end{abstract}

\vspace*{1cm}
\section{Introduction}\label{sec:intro}
The gluon condensate
\begin{equation}
G_2=\langle 0| \frac{\alpha_s}{\pi} G^a_{\mu \nu} G^{a, \mu \nu} |0 \rangle \,\,\, , \label{gc}
\end{equation}
with $G^a_{\mu \nu}$ the gluon field strength tensor,
was introduced in QCD in the framework of the short-distance operator product expansion applied to the two-point correlation function of heavy and light quark current operators \cite{Shifman:1978bx}. It represents the vacuum matrix element of the lowest dimensional gauge-invariant operator constructed by gluon fields,
and parametrizes the long-wavelength fluctuations of the color fields in the nonperturbative QCD vacuum. It typically appears in QCD sum rule analyses, and its value has been determined in a phenomenological way, mainly on the basis of information on the spectrum of heavy quarkonium. Estimates have also been obtained using the dilute instanton-gas approximation \cite{dilute}; the favoured numerical value $G_2 \simeq 0.012$ GeV$^4$ is affected by a large uncertainty \cite{Shifman:1978bx,reviews,narison2012}.

The gluon condensate is related to the QCD trace anomaly which, for massless quarks, reads
\begin{equation}
\Theta^\mu_\mu= \frac{\beta(\lambda)}{\lambda} G^a_{\mu \nu} G^{a, \mu \nu}\,, \label{tranomaly}
\end{equation}
$\Theta^\mu_\mu$ being the trace of the QCD energy momentum tensor, $\displaystyle \lambda=N_c\frac{ g_s^2}{4 \pi}$ the 't Hooft coupling ($N_c$ is the number of colors), and $\beta(\lambda)$ the $\beta$-function of QCD. Hence, the vacuum value of the trace of the QCD energy momentum tensor is connected to (\ref{gc}).

Determinations of the gluon condensate  can be obtained in lattice QCD \cite{Di Giacomo:1981wt}. In this case, the condensate is derived from small Wilson loops, after subtracting a perturbative tail in the lattice coupling constant expansion, whose first coefficients are either analytically computed or fitted to the numerical results. The Wilson loop method can be easily extended to finite temperature.
In particular, simulations in full QCD show that the temperature dependence of the gluon condensate across the deconfinement transition is different for the (chromo) magnetic and electric components: the magnetic condensate is quite independent of T, while the electric condensate decreases as temperature increases
\cite{D'Elia:2002ck}. Numerical information about the chemical potential dependence is not available, at present.

Hence, the gluon condensate reflects relevant features of the strongly coupled color fields in the QCD vacuum. Although it cannot be identified as an order parameter in any QCD phase transition, it is an important quantity to examine when temperature and baryon density are changed. The holographic approach is a suitable method for such a monitoring, in particular  in
models in which the behavior of the quark condensate vs temperature and baryon density can also be studied \cite{Colangelo:2011sr}. In these models, the QCD transition between a chirally symmetric phase and a phase with broken chiral symmetry can be analyzed in the same framework as the deconfinement transition. In lattice simulations,
 at vanishing chemical potential, the two transitions occur close to each other. One can investigate whether
at finite density the two transitions still coincide. The holographic determination of the gluon condensate is useful to gain information about these aspects of QCD. \\
This is the aim of the present study. We analyze the condensate (\ref{gc}) in a holographic model of QCD
described in section \ref{sec:model}. In section \ref{sec:freen} we use the free energy to determine the gluon condensate dependence on temperature and baryon density, while
in section \ref{sec:wloop} we use the small temporal and spatial Wilson loops to study the low-$T$ behaviour of the (chromo) electric and magnetic contributions to the gluon condensate. The conclusions are collected in the last section.

\section{Holographic model}\label{sec:model}
The problem of studying QCD at finite temperature and baryon density can be faced by methods inspired by the gauge/gravity correspondence \cite{Maldacena:1997re} and developed in top-down or bottom-up procedures.
Such approaches aim at investigating the nonperturbative regime of QCD through its possible semiclassical, weakly coupled, higher dimensional dual theory, following the spirit of
the correspondence between the strong-coupling regime of $\CMcal{N}$=4 Super Yang-Mills (SYM) gauge theory
 in a $4d$ Minkowski space and the weak-coupling regime of type IIB string theory in a $5d$ anti-de Sitter (AdS) space, times a compact $5d$ manifold.
In these approaches, the same rules relating operators of the boundary gauge theory to their dual fields are followed \cite{Gubser:1998bc,Witten:1998qj}. Modifications with respect to the AdS/CFT correspondence are introduced, in order to adapt the conjecture to QCD, in particular as far as breaking of scale invariance is concerned.
Far from identifying a unique QCD dual, they lead to the formulation of several phenomenological models in which a few key features of strong interaction phenomenology are encoded.

 Investigations of the phase diagram of QCD, when temperature and density of the hadron system are changed, have recently appeared in this framework \cite{Colangelo:2011sr,Sin:2007ze,Lee:2009bya,Colangelo:2010pe}, with focus on in-medium behavior of hadron properties \cite{Jo:2009xr,Colangelo:2012jy}, as well as on thermodynamics
 \cite{Andreev:2010bv,Stoffers:2010sp}.
Here, we are interested in studying the gluon condensate at increasing temperature and baryon density. We adopt the holographic soft-wall model, formulated to study hadron properties, which uses the occurrence of Regge trajectories in the low-lying hadronic spectra as a guiding information \cite{Karch:2006pv}. The model
 is characterized by a dilaton-like term in the higher dimensional dual theory, introduced to break conformal invariance, and
it has been used to study several aspects of QCD \cite{Colangelo:2007pt,review}.  A good description of known phenomenological features has been achieved in spite of the simplicity of the model.
We follow two ways to determine the gluon condensate, through the free energy and by computing small Wilson loops.

\subsection{Geometry}
Temperature and chemical potential effects can be included in the holographic description by introducing in the  $5d$ AdS space a charged black-hole.
Such a geometry is known as AdS/Reissner-Nordstr\"om (RN) and is characterized, in the Euclidean space, by the metric
%\cite{Wald:1984rg}
\begin{equation}\label{eq:EinsLineEl}
 ds^2=\frac{R^2 e^{2A(z)}}{z^2} \left(f(z)d\tau^2+d\bar x^2+\frac{dz^2}{f(z)} \right) \,\,\, ,
\end{equation}
with coordinates $(\tau, x_1, x_2, x_3, z)$, positive holographic coordinate $z$, $A(z)=0$, and
\begin{equation}\label{eq:fBH}
 f(z)=1-\left( \frac{1}{z_h^4}+q^2 z_h^2 \right) z^4+q^2 z^6 \,\,\, .
\end{equation}
$R$ is the radius of the AdS space, $q$ the charge of the black hole, $z_h$ the position of the black-hole horizon, defined by the condizion $f(z_h)=0$; from now on we will set $R=1$.

In the literature two different modifications have been introduced in the phenomenological set up now known as soft-wall model~\cite{Karch:2006pv}, in order to introduce a mass scale in the theory thus making the $4d$ boundary theory more similar to QCD. 
One choice, proposed in \cite{Karch:2006pv}, consists in including in the action a factor $e^{-\phi(z)}=e^{a_E c^2 z^2}$, while $A(z)=0$ in \eqref{eq:EinsLineEl}. In this framework, thermodynamic properties have been first studied in \cite{Herzog:2006ra}, by calculating the free energy from the gravity action.
The other choice has been considered in \cite{Andreev:2006vy}, and consists in modifying the metric by putting $A(z)=a_E c^2 z^2$ in \eqref{eq:EinsLineEl}, while $\phi(z)=0$; some thermodynamic properties have been investigated in \cite{Andreev:2007zv}. 
$c$ is a dimensionful parameter that breaks conformal symmetry in vacuum ($c\sim \Lambda_{QCD}$).
In this work we focus on the first case, and generalize the study of the thermodynamic properties to the case of nonzero temperature and chemical potential.

$a_E$ is a coefficient that will be fixed hereinafter.
The condition $a_E<0$ is needed to avoid a massless pole in the two-point correlation function of quark vector currents \cite{dilaton}; we shall find the same condition using considerations on the pressure.
As in other bottom-up holographic models of QCD,
the constant $c$ can be fixed from the spectrum of the $\rho$ mesons, which, in both versions of the soft-wall model, is given by $m_n^2=-4 a_E c^2 (n+1)$.

The temperature $T$ is defined by the relation
\begin{equation}\label{eq:temperature}
 T=\frac{1}{4\pi}\left| \frac{df}{dz} \right|_{z=z_h}=\frac{1}{\pi z_h} \left( 1-\frac{q^2 z_h^6}{2} \right)=\frac{1}{\pi z_h} \left( 1-\frac{Q^2}{2} \right)\,,
\end{equation}
where $Q=q z_h^3$.
The temporal component of a U(1) gauge field in the bulk, $A_0(z)$, is dual to the quark number operator $\psi^\dagger \psi$ appearing in the QCD generating functional at finite density.
Following the AdS/CFT dictionary, the boundary value $A_0(0)$ can be related to the source of this operator, i.e. the quark chemical potential: $A_0(0)=i\mu$ (the imaginary unit arises considering the Euclidean spacetime).
The equation of motion for $A_0(z)$ can be obtained from the Maxwell part of the dual $5d$ action
\begin{eqnarray}\label{eq:actionA0}
 S &\propto& \int d^5x \, \sqrt{g}~ e ^{-\phi} \, F_{MN}F^{MN} \,,
\end{eqnarray}
with $F_{MN}=\partial_M A_N-\partial_N A_M$. 
The general solution for the only non-vanishing component $A_0(z)$ involves two coefficients: $A_0(z)=i ( B_1 - \frac{B_2}{2 a_E c^2} e^{-a_E c^2 z^2})$.
Imposing $A_0(0)=i \mu$, together with the condition that, for $a_E\to 0$, the solution coincides with the RN one, $A_0^{RN}(z)=i(\mu-\sqrt{3 g_5^2} q z^2)\,\,\,$\footnote{In the AdS/RN solution the coefficient $\sqrt{3 g_5^2}q$ is fixed solving the Einstein equation $f''-3 f'/z=\frac{1}{g_5^2} z^2 A_0'^2$ together with the equation of motion $A_0''-A_0'/z=0$, and imposing that $f$ has the expression \eqref{eq:fBH}. Notice that in \cite{Andreev:2010bv} the coefficient is $\sqrt{3}/2$ since $g_5^2=1/4$.}, we get:
\begin{equation}\label{eq:solA0}
 A_0(z)=i \left(\mu-\frac{ \sqrt{3 g_5^2}\, q}{a_E\, c^2} \left(1-e^{-a_E c^2 z^2}\right) \right)\,.
\end{equation}
$A_0(z)$ gets the same expression also in the second version of the soft-wall model.
The vanishing of $A_0(z)$ at the horizon, $A_0(z_h)=0$, sets a relation between the chemical potential and the charge of the black hole:
\begin{equation}\label{eq:mu}
 \mu=\frac{\sqrt{3 g_5^2} \, q}{ a_E\, c^2} \left( 1-e^{-a_E c^2 z_h^2} \right) =\frac{\sqrt{3 g_5^2} \, Q}{a_E \, c^2 \, z_h^3} \left( 1-e^{-a_E c^2 z_h^2} \right) \,.
\end{equation}

In the following we compute the gluon condensate and a few thermodynamical quantities, comparing the results to other determinations.\footnote{
A computation of thermodynamical quantities has been carried out in a holographic framework with a different metric \cite{He:2013qq}.} Indeed,
in the confined phase, modifications of various observables with respect to $T,\mu=0$ are expected \cite{Cohen:1991nk}.
Our model exhibits a non-trivial structure in the low-temperature/finite-density region, which is discussed in the following section. 
This structure is a consequence of the ansatz for the black-hole function $f(z)$ in \eqref{eq:fBH}, and is different from the case of a dynamically determined $f(z)$ \cite{Li:2011hp}. 
Finally, let us remark that
 the confined phase could be holographically described by a different metric, Thermal AdS, without black holes, and a Hawking-Page (HP) transition between Thermal AdS and the black-hole metric could occur, associated to the deconfinement transition in QCD \cite{Herzog:2006ra}. In the following we do not consider such a possibility.

\subsection{Low-temperature}
\begin{figure}[h!]
 \centering
 \includegraphics[width=8cm]{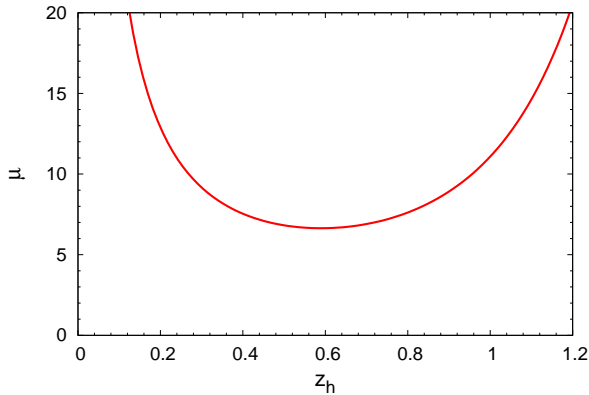}
\caption{$\mu(z_h)$ from Eq.~\eqref{eq:mu}, setting $Q=\sqrt{2}$, $g_5^2=1$, $c=1$ and $a_E\simeq-2.5$.}
 \label{fig:Qsqrt2}
\end{figure}

The low-temperature regime described by Eqs.~\eqref{eq:temperature},\eqref{eq:mu} deserves a detailed discussion.
From Eq.~\eqref{eq:temperature} one sees that $T=0$ corresponds either to $Q=\sqrt{2}$ or to $z_h\to \infty$.
If $Q=\sqrt{2}$, only some values of $\mu$ can be obtained varying $z_h$, since $\mu(z_h)$ has a positive minimum for any $a_E<0$, as shown in Fig.~\ref{fig:Qsqrt2}.
Therefore, lower values of the chemical potential can be reached only from high values of $z_h$ and very small values of $Q$ ($Q$ should be small enough to contrast the exponential divergence $e^{z_h^2}$ in $\mu$).
If we fix $T$, take $z_h$ from \eqref{eq:temperature} and substitute in \eqref{eq:mu}, we can represent $\mu$ as a function of the charge $Q$ at fixed temperature:
\begin{equation}\label{eq:mufixedT}
 \mu(Q)=\frac{\sqrt{3 g_5^2} Q \pi^3 T^3}{a_E (1-Q^2/2)^3} \left( 1- e^{-a_E (1-Q^2/2)^2/(\pi T)^2} \right)\,.
\end{equation}

\begin{figure}[h!]
 \centering
 \includegraphics[width=8cm]{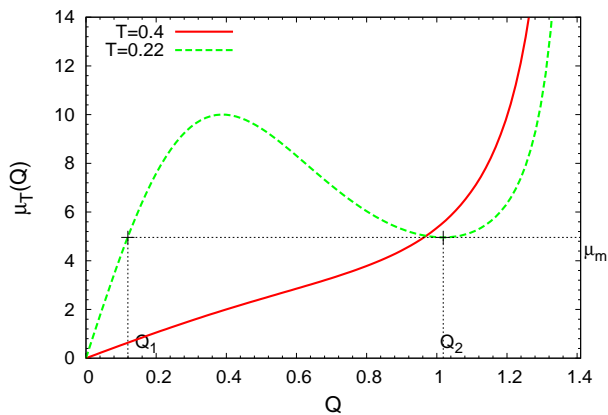}
\caption{Chemical potential $\mu$ versus $Q$ for two values of temperature, $T=0.4$ (plain line) and $0.22$ (dashed line), with parameters $c$, $g_5^2$ and $a_E$
 as in Fig.\ref{fig:Qsqrt2}. For the lowest temperature the relation $\mu_T$ vs $Q$ is not one-to-one. }
 \label{fig:muvsQ}
\end{figure}

\begin{figure}[h!]
 \centering
\includegraphics[width=8cm]{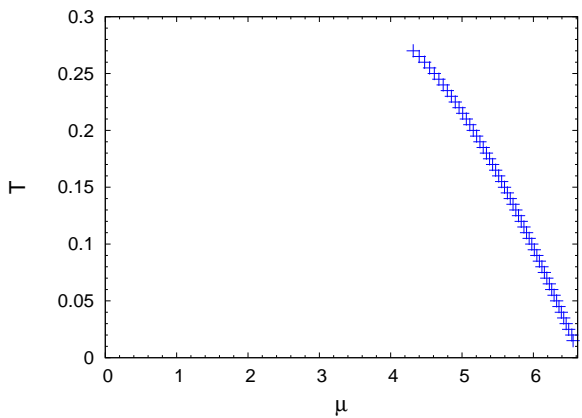}
\caption{Points in the plane $(T,\mu)$ corresponding to a jump, as in Fig.\ref{fig:muvsQ}, and where the thermodynamical quantities present a discontinuity. The values of $c$, $g_5^2$ and $a_E$
are the same as in Fig.\ref{fig:Qsqrt2}.}
 \label{fig:jump}
\end{figure}

In Fig.~\ref{fig:muvsQ} $\mu_T(Q)$ is plotted for two values of temperature, $T=0.4$ and $T=0.22$ (in units of $c$, with $g_5^2=1$ and putting $a_E\sim -2.5$ as it will be set in the next section).
For the higher temperature, $T=0.4$, there is a one-to-one correspondence between $Q$ and $\mu$, while at $T=0.22$ it is not possible to continuously obtain lower values of $\mu$ lowering $Q$, due to the presence of a local minimum; at some point, decreasing $\mu$ there is a jump from one value of $Q$ to another one.
In the plane ($\mu,T$) the values for which there is this jump are depicted in Fig.~\ref{fig:jump} (for the same values of  $c$, $g_5^2$ and $a_E$).
A different value of $c$ would rescale both $T$ and $\mu$, while a different value of $g_5^2$ would only affect the scale of $\mu$ \cite{Colangelo:2010pe}.
The jump is due to the form of $f(z)$ in Eq.~\eqref{eq:fBH}, and disappears once the Einstein equations for a theory with modified RN metric with dilaton are solved.
In our model, a first-order phase transition occurs at high density and low temperature, characterized by a discontinuity in all thermodynamical quantities.

\section{Gluon condensate from the free energy}\label{sec:freen}
One of the methods we use to compute the gluon condensate is based on the relation \eqref{tranomaly} and involves the computation of the trace of the energy-momentum tensor through thermodynamical functions.
We make use of the AdS/CFT correspondence relation
\begin{equation}
\CMcal{Z}\sim e^{-S}
\end{equation}
between the $4d$ gauge partition function $\CMcal{Z}$ and the $5d$ gravity action $S$ to compute the free-energy density
\begin{equation}
\CMcal{F}=-\frac{T}{V}\log \CMcal{Z}\,.
\end{equation}
In the RN model the free-energy density gets two contributions, from the Einstein-Hilbert and  the Maxwell terms of the action:
\begin{eqnarray}\label{eq:freeen}
 \CMcal{F} &=& -\frac{1}{16 \pi G_N}\int_0^{z_h} dz\, \sqrt{g} \left( \CMcal{R}-2\Lambda-\frac{1}{4g_5^2} F^2\right) \nonumber \\
 &=& -\frac{1}{16 \pi G_N}\int_0^{z_h} dz\, \frac{1}{z^5} \left( -8 -2 Q^2 \frac{z^6}{z_h^6} -\frac{1}{2g_5^2} z^4 A_0'(z)^2\right) \,;
\end{eqnarray}
$G_N$ is the Newton constant in $5d$.
In the previous sections we have introduced two possible modifications of the RN model.
The free energy of the model with $A(z)=0$ and $\phi=-a_Ec^2z^2$, assuming a non-dynamical dilaton, reads
\begin{eqnarray}\label{eq:freeen}
 \CMcal{F} &=& -\frac{1}{16 \pi G_N}\int_0^{z_h} dz\, \frac{e^{ a_E c^2 z^2}}{z^5} \left( -8 -2 Q^2 \frac{z^6}{z_h^6} -\frac{1}{4g_5^2} F^2\right) \nonumber \\
 &=& -\frac{1}{16 \pi G_N}\int_0^{z_h} dz\, \frac{e^{ a_E c^2 z^2}}{z^5} \left( -8 -2 Q^2 \frac{z^6}{z_h^6} -\frac{1}{2g_5^2} z^4 A_0'(z)^2\right) \nonumber \\
&=& \frac{1}{8 \pi G_N} (\CMcal{F}_1+\CMcal{F}_2) \,.
\end{eqnarray}

Eq.~\eqref{eq:freeen} needs to be regularized. To this aim, we write $\CMcal{F}_1$ as
\begin{eqnarray}\label{eq:f1}
 \CMcal{F}_1(z_h) &=& 4 \int_0^{z_h} dz\, \frac{e^{a_E c^2 z^2}}{z^5} \nonumber\\
 &=& 4 \int_0^{z_h} dz \left[\frac{e^{a_E c^2 z^2}}{z^5} - \frac{1}{z^5} - \frac{a_E c^2}{z^3} -\frac{ a_E^2 c^4}{2z} \right]\nonumber\\
 && + 4 \int_\epsilon^{z_h} dz \left[ \frac{1}{z^5} + \frac{ a_E c^2}{z^3} + \frac{ a_E^2 c^4}{2z} \right]_{\epsilon\to 0}\,,
 \end{eqnarray}
 obtaining
\begin{equation}\label{eq:subtr}
\CMcal{F}_1^{REG}(z_h) = -\frac{e^{ a_E c^2 z_h^2}}{z_h^4} -\frac{1}{2} a_E^2 c^4\left( -3 +2 \gamma_E-2 \Gamma[-1,- a_E c^2 z_h^2] + \log[ a_E^2] \right)\,. \\
\end{equation}
 This regularization scheme, consisting in the subtraction of the divergent terms $1/\epsilon^4$, $2 a_E c^2/\epsilon^2$ and $- a_E^2 c^4 \log[c^2\epsilon^2]$,
is chosen in order to obtain a vanishing pressure at zero temperature.
Eq. \eqref{eq:subtr} shows that $a_E<0$ is required.
On the other hand, using Eq.~\eqref{eq:solA0}, $\CMcal{F}_2$ reads:
 \begin{eqnarray}\label{eq:f2}
 \CMcal{F}_2(z_h, Q) &=& \frac{Q^2}{z_h^6} \int_0^{z_h} dz\, z\, e^{ a_E c^2 z^2} + \frac{1}{4g_5^2} \int_0^{z_h} dz \frac{e^{a_E c^2 z^2}}{z} A_0'(z)^2 \nonumber\\
&=& \frac{Q^2}{2 a_E\, c^2\, z_h^6} \left(e^{a_E c^2 z_h^2}-1\right)-\frac{3 Q^2}{2 a_E \, c^2\, z_h^6} \left( 1-e^{-a_E\, c^2\, z_h^2}\right) \,.
\end{eqnarray}
The parameters $z_h$ and $Q$ are related to $T$ and $\mu$ through Eqs.~\eqref{eq:temperature} and \eqref{eq:mu}.
For values of $T$ for which the relation $\mu_T(Q)$ in Eq.~\eqref{eq:mufixedT} is not one-to-one, we choose the solution shown in Fig.~\ref{fig:muvsQ}: if $\mu$ is greater than the relative minimum ($\mu>\mu_m$), the charge $Q$ is taken in the range $Q_2<Q<\sqrt{2}$, while if $\mu<\mu_m$ we take $0<Q<Q_1$.
This choice corresponds to a lower free energy.
In the numerical analysis we set $c=1$ and $g_5^2=1$.

\subsection{Thermodynamical quantities}
The pressure
\begin{equation}
p=T \frac{\partial \ln \CMcal{Z}}{\partial V},
\end{equation}
for large homogeneous systems, is related to the free energy density
\begin{equation}
 p=-\CMcal{F}\,. \label{press}
\end{equation}
The entropy density can be computed using
\begin{equation}\label{eq:entropy}
 s=\frac{\partial \left[T \log \CMcal{Z}\right]}{\partial T}=\frac{\partial p}{\partial T}\,.
\end{equation}
In the limit $(\mu ,T)\to 0$, the pressure obtained from \eqref{eq:subtr},\eqref{eq:f2} behaves as
\begin{equation}
p(T,\mu) \to \frac{1}{8\pi G_N} \frac{1}{2} a_E^2\, c^4\, (2 \gamma_E-3+2 \log(- a_E)) \,\,\, \label{press1}
\end{equation}
and vanishes if $a_E=-e^{3/2-\gamma_E}\sim -2.5$.
This condition allows us to set the value of $a_E$.\footnote{In presence of a Hawking-Page phase transition this condition on $a_E$ is not required.}
On the other hand,  the $T\to \infty$ limit of the pressure sets the coefficient in front of the free energy.
Indeed, comparing
\begin{equation}
p \to \frac{1}{8 \pi G_N} (\pi^4 T^4 + ...)
\end{equation}
to the result for a free massless gas of bosons in thermal equilibrium \cite{Chodos:1974je,Andreev:2007zv}, we find $1/8\pi G_N=8/(45 \pi^2)$ in SU(3) pure gauge theory.

To infer how the pressure changes at different values of the chemical potential $\mu$, it is convenient to look at the ratio $p/T^4$. Notice that all the dimensionful quantities can ge given in units of $c$. Using the value of $a_E$ found before and $m_0=m_\rho=0.776$GeV, one gets $c\simeq0.25$GeV.

\begin{figure}[t!]
 \centering
 \includegraphics[width=8cm]{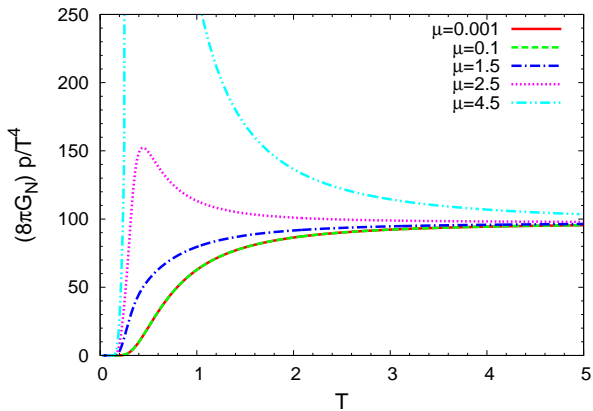}
\caption{Ratio $p(T,\mu)/T^4$ versus $T$, for chemical potential $\mu=0.001$ (plain red line), $0.1$ (dashed green line), $1.5$ (dot-dashed blue line), $2.5$ (dotted purple line) and $4.5$ (dot-dot-dashed cyan line). $T$ and $\mu$ are in units of $c$; $g_5^2=1$.}
\label{fig:pressure}
\end{figure}

\begin{figure}[h!]
 \centering
 \includegraphics[width=8cm]{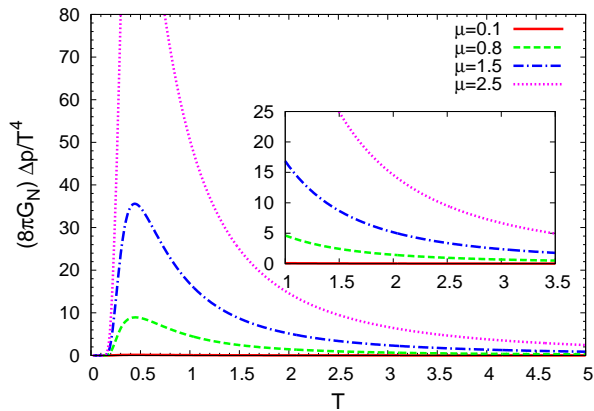}
 \caption{$\Delta p/T^4$ versus $T$. In the inset, the region of temperature which follows the peaks is enlarged. The plain red line corresponds to $\mu=0.1$, the dashed green line to $\mu=0.8$, the dot-dashed blue line to $\mu=1.5$, the dotted purple line to $\mu=2.5$. $T$ and $\mu$ are in units of $c$; $g_5^2=1$.}
 \label{fig:Deltap}
\end{figure}

\begin{figure}[h!]
 \centering
 \includegraphics[width=8cm]{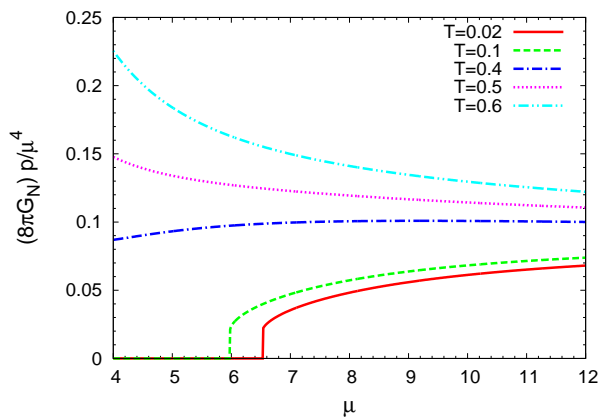}
 \caption{$p/\mu^4$ versus the chemical potential $\mu$ for several values of temperature: $T=0.02$ (plain red line), $0.1$ (dashed green line), $0.4$ (dot-dashed blue line), $0.5$ (dotted purple line) and $0.6$ (dot-dot-dashed cyan line). $T$ and $\mu$ are in units of $c$; $g_5^2=1$.}
 \label{fig:pdivmu}
\end{figure}

The ratio $p/T^4$ is shown in Fig.~\ref{fig:pressure}.
For small values of the chemical potential, $p/T^4$ has a monotonic $T$ dependence, and saturates at $T/c \geq 3$, a result common to other approaches. At higher values of $\mu$ the asymptotic value at $T\to\infty$ (independent of $\mu$) is reached from above.
The figure shows that, as the chemical potential increases, $p/T^4$ reaches a peak whose position coincides with the points in Fig.~\eqref{fig:jump} for $\mu>4$, and vanishes for $T\to 0$.
This behavior is different from the one found, e.g., in Fig.~4.10 of \cite{Vuorinen:2004rd}, where the low-temperature, high-density region has been scrutinized using perturbation theory.
The variation of the pressure with respect to its value at $\mu=0$ can be observed in Fig.~\ref{fig:Deltap}, where the quantity $\Delta p/T^4=(p(T,\mu)-p(T,0))/T^4$ is plotted versus $T$.
Similar results have been obtained in \cite{Allton:2003vx,Gavai:2003mf,Aoki:2008rt,Borsanyi:2012cr} (see, in particular,  Fig.~6(b), Fig.~2, Fig.~5 and Fig.~1 (left panel) of these references, respectively) and through perturbation theory, as shown in Fig.~2 (left panel) of Ref.~\cite{Vuorinen:2003fs}, in Fig.~5 of Ref.~\cite{Haque:2012my}, and in Fig.~7 of Ref.~\cite{Mogliacci:2013mca}.
The ratio $p/\mu^4$ as a function of the chemical potential, Fig.~\ref{fig:pdivmu}, shows the same effect observed in Fig.~\ref{fig:pressure}: as the temperature increases, the curve approaches the asymptotic value from above.

\begin{figure}[h!]
 \centering
\includegraphics[width=8cm]{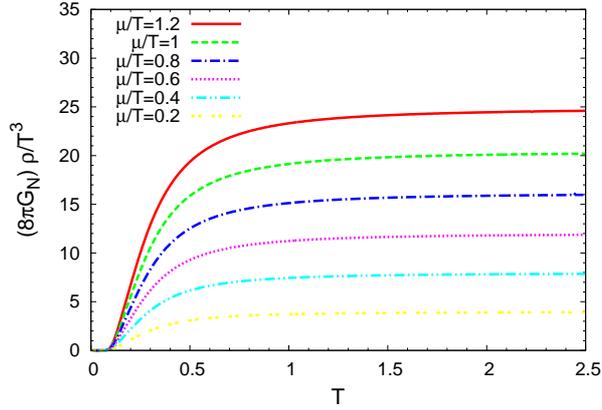}
 \caption{$\rho/T^3$ versus $T$ for several values of the ratio $\mu/T$: $\mu/T=1.2$ (plain red line), $1$ (dashed green line), $0.8$ (dot-dashed blue line), $0.6$ (dotted purple line), $0.4$ (dot-dot-dashed cyan line) and $0.2$ (dot-dot yellow line). $T$ and $\mu$ are in units of $c$; $g_5^2=1$.}
 \label{fig:density}
\end{figure}

The quark density $\rho=\partial p/\partial \mu$ is plotted versus $T$ in Fig.~\ref{fig:density} for several values of $\mu/T$.
The ratio $\rho/T^3$ increases near the critical temperature, with a slope increasing with $\mu/T$. At $\mu=0$ this behavior reproduces the one found in lattice QCD (see Figs.~5-6 of Ref.~\cite{Aoki:2008rt}).

\begin{figure}[bh!]
 \centering
 \includegraphics[width=8cm]{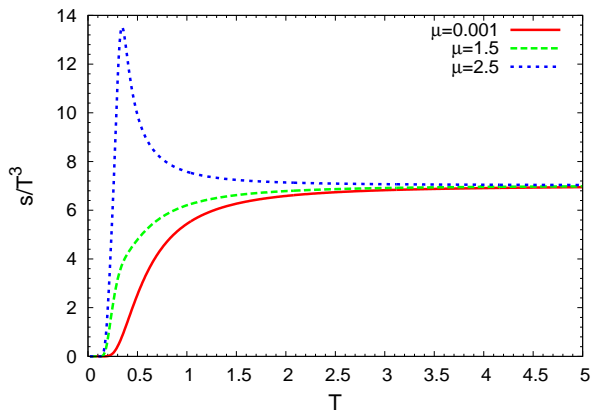}
\caption{Entropy density, divided by $T^3$, computed from Eq.~\eqref{eq:entropy}, versus temperature $T$, for some values of the chemical potential.}
\label{fig:entropy}
\end{figure}

The last thermodynamical quantity needed to get the gluon condensate is the entropy density, which
 can be computed by Eq.~\eqref{eq:entropy}.
Looking at Fig.~\ref{fig:entropy} it is possible to appreciate a property of the holographic model introduced here, the vanishing of the entropy as $T\to 0$ (for $\mu\lesssim 6.8$).
This property is not shared by the RN model \cite{Ammon:2011hz} (unless a Hawking-Page transition occurs
for high $\mu$ and low temperature to another phase described by a different metric).

\subsection{$T$ and $\mu$ dependence of the gluon condensate}

\begin{figure}[h]
\begin{center}
\includegraphics[width=7cm]{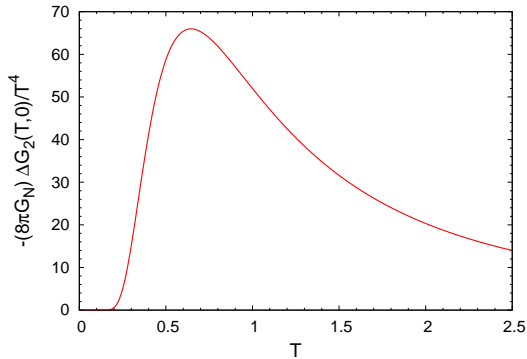}
\caption{$- \Delta G_2(T,\mu)/T^4$ as a function of $T$ at $\mu=0$.
}
\label{fig:G2m0}
\end{center}
\end{figure}

\begin{figure}[h!]
 \centering
 \includegraphics[width=8cm]{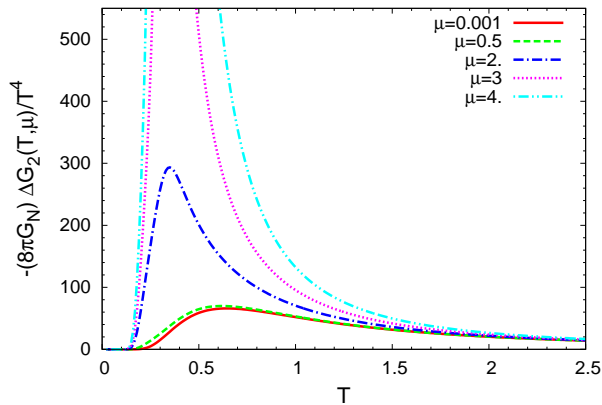}
\caption{$- \Delta G_2(T,\mu)$ versus $T$ for several values of the chemical potential: $\mu=0.001$ (plain red line), $\mu=0.5$ (dashed green line), $\mu=2$ (dot-dashed blue line), $\mu=3$ (dotted purple line), $\mu=4$ (dot-dot-dashed cyan line). $T$ and $\mu$ are in units of $c$; $g_5^2=1$. }\label{fig:G2}
\end{figure}

The variation of the gluon condensate versus temperature and density can be obtained from the energy density, using the relation $\epsilon=Ts-p+\mu\rho$, together with the formula
\begin{equation}
 \Delta G_2(T,\mu)=G_2(T,\mu)-G_2(0,0)= -\epsilon(T,\mu)+ 3 p(T,\mu)\,
\end{equation}
derived, e.g., in \cite{Leutwyler:1992cd} at finite temperature; contributions from additional degrees of freedom are discussed in \cite{Castorina:2007qv}, while the condensate in nuclear matter is studied, e.g., in \cite{Baldo:2003id}. We make use of the relation
\begin{equation}
\Delta G_2(T,\mu)=4 p(T,\mu)-T s(T,\mu)-\mu \rho(T,\mu) \, .
\end{equation}
In Fig.~\ref{fig:G2m0} we plot the $T$-dependence of $- \Delta G_2(T,\mu)/T^4$ at $\mu$=0, which reproduces the shape obtained by lattice QCD (see, e.g., Fig.~2(b) of Ref.~\cite{Boyd:1995zg}). The difference is the smaller slope in reaching the maximum value.
The agreement between holographic and lattice QCD results is noticeable, and makes us confident on the reliability of the results in other regions, namely at finite $\mu$.
The quantity $- \Delta G_2(T,\mu)/T^4$ at finite density is depicted in Fig.~\ref{fig:G2}. Peaks are found in correspondence of the points in Fig.~\ref{fig:jump} (for $\mu >4$);
 the height of each peak increses with the chemical potential. For high values of $\mu$, $\Delta G_2$ becomes a monotonic function of temperature, while at high temperatures it becomes independent of $\mu$: asymptotically, $\Delta G_2$ behaves as in the limit of large number of colors, in which no density dependence is expected.
The first order phase transition manifests by a divergence at a critical low-temperature.

\section{Gluon condensate from small Wilson loops: (chromo) electric and magnetic contributions}\label{sec:wloop}

The gluon condensate can also be computed in a different way, expanding the vacuum expectation value of a small Euclidean Wilson loop $W(\CMcal{C})$ in powers of the area $s$ of the loop.
The method is similar to the one adopted in lattice QCD to compute $G_2$  \cite{Di Giacomo:1981wt}.
The expansion can be written as
\begin{equation}\label{eq:logW}
\log \left( \langle W \rangle \right) =-\sum_n c_n \alpha_s^n-\frac{\pi^2}{36}Z G_2 s^2 +\CMcal{O}(s^3)\,
\end{equation}
and involves a perturbative series in $\alpha_s$; the gluon condensate $G_2$ appears in the coefficient of the $\CMcal{O}(s^2)$ term.
 $Z$ is a renormalization constant that we set to $Z$=1, following \cite{Andreev:2007vn}.

We compute $\log \left( \langle W \rangle \right)$ at $ \mu=0$ and small values of $T$  in Eq. \eqref{eq:logW} in the holographic approach, extending the calculations
 at $T=0$ made in \cite{Andreev:2007vn,Goity:2012yj}.
We consider a small circular Wilson loop $\CMcal{C}$ of radius $a$, whose expectation value can be computed through the Nambu-Goto action, i.e. determining the minimal area of the worldsheet spanned by a string in the $5d$ bulk with endpoints attached to $\CMcal{C}$, according to the gauge/gravity duality prescriptions:
\begin{equation}
\langle W(\CMcal{C}) \rangle \sim e^{-S_{NG}}\,.
\end{equation}
The Nambu-Goto action is
\begin{equation}\label{eq:NGaction}
S_{NG}=\frac{1}{2\pi \alpha^\prime} \int d^2\xi \, \sqrt{\gamma}\,,
\end{equation}
with $(\xi_1,\xi_2)$ the worldsheet coordinates and $\gamma$ the induced metric.
We choose $\xi_1=r$ and $\xi_2=\phi$, where ($r,\phi$) is the representation in polar coordinates of $(x,\tau)$ in the case of a temporal Wilson loop, and $(x,y)$ for a spatial one.
Notice that at finite temperature the expectation values of a temporal and a spatial Wilson loop do not coincide, and can be related to different quantities, the (chromo) electric and magnetic component of the gluon condensate, respectively \cite{Adami:1990sv}.

To compute these quantities, we follow \cite{Andreev:2007vn} and use the line element
\begin{equation}
ds^2=\frac{e^{c_S^2 z^2}}{z^2} \left( f(z) d\tau^2 + d\bar x^2+\frac{dz^2}{f(z)}\right) \,\,\, ,
\end{equation}
where $e^{c_S^2 z^2}/z^2$ is  the warp factor.
This factor generates an area law for the quark-antiquark static potential at $T=0$ \cite{Andreev:2006ct}. The numerical value of the scale $c_S$ has been fixed to $c_S=0.67$ GeV from the $\rho$ meson spectrum, as in \cite{Andreev:2007vn}.

We redefine $t=r/a$ and $\psi=z^2/a^2$, and introduce the dimensionless parameter $\lambda=a^2 c_S^2$ \footnote{A different choice of the parameter $\lambda$, e.g. $\lambda=a^2 T^2$, gives the same result for the gluon condensate.}.
The circular loop is centered at $(0,0)$ with radius $a$, and the coordinates are limited by $0\leqslant x,y,\tau \leqslant a$.
For the temporal loop, it must be $\tau<1/T$ for regularity of the metric \cite{Witten:1998zw}; this can be achieved if $a<1/T$.
Including in \eqref{eq:NGaction} the induced metric and integrating in the angle $\phi$, the Nambu-Goto action for a spatial Wilson loop reads:
\begin{eqnarray}
S_{NG}^{y} &=&\int_0^a dr \, \CMcal{L}^{y}(\psi,\psi',r) \nonumber\\
&=& \int_0^1 dt \, t\, \frac{e^{\lambda \psi}}{\psi}\, \sqrt{1+\frac{1-t^2}{4t^2}\frac{\psi'^2}{\psi} \frac{1}{1-\lambda^2\pi^4 \frac{T^4}{c_S^4} \psi^2}} \,,
\end{eqnarray}
while for a the temporal Wilson loop it is given by
\begin{eqnarray}
S_{NG}^{\tau}&=&\int_0^a dr \, \CMcal{L}^{\tau}(\psi,\psi',r) \nonumber\\
&=& \frac{1}{\pi} \int_0^1 dt \, t\, \frac{e^{\lambda \psi}}{\psi}\, \left(\sqrt{A} \mbox{ E}(-B/A) + \sqrt{A+B} \mbox{ E}(B/A+B)\right)\, .
\end{eqnarray}
 E$(x)$ is the complete elliptic integral of the second kind, and
\begin{equation}
A=1-\psi^2 T^2/c_S^2 \pi^4 +\frac{1-t^2}{4t^2} \frac{\psi'^2}{\psi} \frac{1}{1-\psi^2\lambda^2 T^2/c_S^2 \pi^4} \,, \qquad B=\frac{1-t^2}{4t^2} \frac{\psi'^2}{\psi} \frac{-\psi^2\lambda^2 T^2/c_S^2 \pi^4}{1-\psi^2\lambda^2 T^2/c_S^2 \pi^4} \,.
\end{equation}
The action and the solution of the equation of motion can be expanded for small $\lambda$:
\begin{eqnarray}
 S_{NG}&=&S_0+\lambda S_1+\lambda^2 S_2 + \CMcal{O}(\lambda^3) \nonumber \\
\psi&=&\psi_0+\lambda \psi_1+\lambda^2 \psi_2 + \CMcal{O}(\lambda^3) \,\,\, ,
\end{eqnarray}
with
\begin{eqnarray}
S_0 &=& -1 \label{eq:SNG0} \\
S_1 &=& \frac{5}{3}\label{eq:SNG1} \\
S_2^{\tau/y} &=& \frac{7}{90} \left(85\mp 2 \pi^4 \frac{T^4}{c_S^4} -120 \log 2\right) \, .
\end{eqnarray}
$S_0$ and $S_1$ have the same expression for the spatial and temporal Wilson loop, while the temporal Nambu-Goto action of $\CMcal{O}(\lambda^2)$, $S_2^\tau$, differs from the spatial $S_2^y$ in the sign of the $T^4$ term.
It is worth noticing that the linear term $S_1$ in Eq.~\eqref{eq:SNG1} does not vanish in the soft-wall model.
As observed from the high $Q^2$ expansion of  two-point correlation functions of quark or gluon currents, e.g. in Ref.~\cite{Colangelo:2007pt}, in the soft-wall model a dimension-two condensate emerges. In QCD no  local gauge-invariant operator of dimension two can be defined; however, the possible existence and meaning of this quantity is still the subject of discussions \cite{narison}.
The gluon condensate can be extracted from the $\lambda^2 S_2$ term, and its electric ($e$) and magnetic ($m$) parts are: 
\begin{equation}\label{eq:G2plaq}
G_2^{e/m}(T) = \frac{14 c_S^4}{5 \pi^4} \left(85\mp 2 \pi^4 \frac{T^4}{c_S^4} -120 \log 2\right) \,.
\end{equation}
\begin{figure}[h!]
 \centering
 \includegraphics[width=8cm]{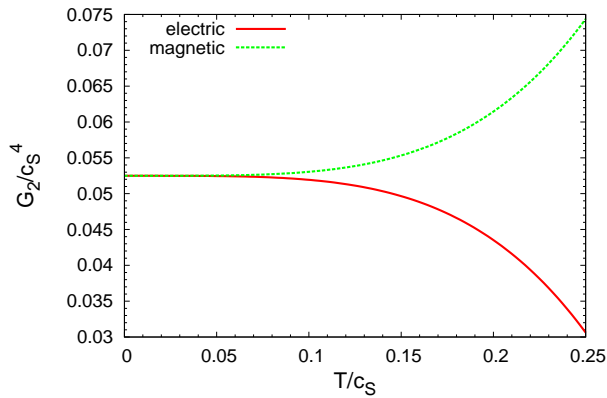}
\caption{(Chromo) electric (continuous red line) and magnetic (dashed green line) component of the gluon condensate $G_2^{e/m}(T)$ for small $T$ and in units of $c_S$, computed from small circular Wilson loops.
}
\label{fig:G2Plaq}
 \end{figure}
The result is depicted in Fig.~\ref{fig:G2Plaq}. The corrections to the magnetic and electric components of the gluon condensate are equal in size but opposite in sign,
and negative for the electric component, therefore the full gluon condensate gets no corrections. This is reminiscent of the result obtained in a perturbative calculation of the smallest Wilson loop, in which the electric and magnetic terms remain equal at $\CMcal{O}(g^2)$ \cite{Boyd:1996bx}.

The result in Eqs. \eqref{eq:SNG0}-\eqref{eq:G2plaq} turns out to be valid at low temperatures, since the coefficient of the second order term grows as $T^4$. Indeed, this is confirmed by a comparison with Fig.~3 of Ref.~\cite{Eletsky:1992rs}, where the dependence on temperature of the electric and magnetic components of the gluon condensate has been computed, finding that the two quantities increase as $T^4$, with the same coefficient and opposite sign.
The approximations involved in the calculation inhibit the extension of  the result to intermediate temperatures.

\section{Conclusions}\label{sec:conclusions}
We have studied the gluon condensate when both temperature and chemical potential are switched on, starting from the free energy of the theory living in a $5d$ space with AdS/Reissner-Nordstr\"om metric with a dilaton-like term in the action.
We have found that the $T-$dependence of the gluon condensate coincides with the one obtained in lattice QCD at $\mu=0$. At large temperature and density, the condensate does not depend on $\mu$, as expected on the basis of large $N_c$ arguments. At low temperature a peak is found, whose height increases with the chemical potential.
Similar results are obtained as well by considering a different modification of the RN model, in which the metric is distorted; differences arise mainly in the numerical value of the parameters $a_E$ and $c$, and so in the scale of the dimensionless quantities $T/c$ and $\mu/c$.
The model gives rise to a peculiar structure of the phase diagram, with a first order phase transition at high values of the chemical potential and low $T$. This is reflected in a discontinuity of the thermodynamical quantities for those values of $T$ and $\mu$,  also visible in the gluon condensate.

For finite temperature and vanishing density, we have also computed $G_2$ from small Wilson loops. This method allows to separate the (chromo) electric and magnetic
components of the condensate, related to temporal and spatial loops, respectively.
This method shows that the two components have the same temperature dependence, but with a different sign, so that the gluon condensate remains unchanged. This is an indication that the method based on small Wilson loops can be properly used only at low temperatures:  the coefficient of the expansion is proportional to $T^4$, and smaller and smaller values of the radius of the loop must be chosen to make the series convergent, and the extension to high temperature is unreliable.
This is also confirmed by a comparison with the outcomes of the first method we have used for computing the gluon condensate. In fact, in Fig. 9 one can notice that the gluon condensate remains constant as well, up to temperatures around 0.2$c$, hence the analysis with the Wilson loop should be reliable up to $T/c_S \sim 0.2 c/c_S \sim 0.08$.
Using a different function $f(x)$ in the black-hole metric, namely the solution of the Einstein equation once the warp factor is fixed, the (chromo) electric and magnetic components of the condensate have the same, but opposite in sign, behavior vs temperature, with a different profile with respect to the one found here, and asymptotic $T^4$ dependence.

 \section*{Acknowledgments}
 We thank Paolo Castorina, Fulvia De Fazio, Massimo Mannarelli and Dario Zappal\`a for discussions.
 This work is partially supported by the Italian Miur PRIN 2009.
 FZ is partially supported by the National Natural Science Foundation of China under Grant No. 11135011.

\end{document}